\documentclass[12pt,aps,prd,preprint,tightenlines,%superscriptaddress,
   showpacs,nofootinbib]{revtex4}
\newcommand{\PRE}[1]{{#1}} % Use if preprint style
\usepackage{bm} \usepackage{epsfig}

\newcommand{\mstar}{M_{*}}

\newcommand{\ifb}{\text{fb}^{-1}} 
\newcommand{\ev}{\text{eV}}
 
\newcommand{\mev}{\text{MeV}}
\newcommand{\gev}{\text{GeV}} 
\newcommand{\tev}{\text{TeV}}

\newcommand{\cm}{\text{cm}} 
\newcommand{\m}{\text{m}}

\newcommand{\g}{\text{g}} 
\newcommand{\s}{\text{s}}
\newcommand{\yr}{\text{yr}} 

\newcommand{\kton}{\text{kton}}

\newcommand{\eqref}[1]{Eq.~(\ref{#1})}

\newcommand{\secref}[1]{Sec.~\ref{sec:#1}}
\newcommand{\secsref}[2]{Secs.~\ref{sec:#1} and \ref{sec:#2}}
\newcommand{\figref}[1]{Fig.~\ref{fig:#1}}

\newcommand{\tableref}[1]{Table~\ref{table:#1}}

\newcommand{\NLSP}{\text{NLSP}}

\newcommand{\mNLSP}{m_{\NLSP}}

\newcommand{\gravitino}{\tilde{G}}

\newcommand{\stau}{\tilde{\tau}}
\newcommand{\slepton}{\tilde{l}}

\newcommand{\mgaugino}{M_{1/2}}

\newcommand{\rin}{r_{\text{in}}}
\newcommand{\rinwe}{r_{\text{in}}^{\text{we}}}
\newcommand{\mwe}{\text{mwe}}
\newcommand{\mcubedwe}{\text{m$^3$we}}

\begin{document}

\preprint{UCI-TR-2004-30}  \preprint{hep-ph/0409278}

\title{
\PRE{\vspace*{1.5in}}
Slepton Trapping at the Large Hadron and \\
International Linear Colliders
\PRE{\vspace*{0.3in}}
}

\author{Jonathan L.~Feng}
\affiliation{Department of Physics and Astronomy, University of
California, Irvine, CA 92697, USA
\PRE{\vspace*{.5in}}
}

\author{Bryan T.~Smith%
\PRE{\vspace*{.2in}}
} 
\affiliation{Department of Physics and Astronomy, University of
California, Irvine, CA 92697, USA
\PRE{\vspace*{.5in}}
}

%\date{September 2004}

\begin{abstract}
\PRE{\vspace*{.3in}} We consider supergravity with a gravitino
lightest supersymmetric particle. The next-to-lightest supersymmetric
particle (NLSP) decays to the gravitino with lifetime naturally in the
range $10^4 - 10^8~\s$.  However, cosmological constraints exclude
lifetimes at the upper end of this range and disfavor neutralinos as
NLSPs, leaving charged sleptons with lifetimes below a year as the
natural NLSP candidates.  Decays to gravitinos may therefore be
observed by trapping slepton NLSPs in water tanks placed outside Large
Hadron Collider (LHC) and International Linear Collider (ILC)
detectors and draining these tanks periodically to underground
reservoirs where slepton decays may be observed in quiet environments.
We consider 0.1, 1, and 10 kton traps and optimize their shape and
placement.  We find that the LHC may trap tens to thousands of
sleptons per year.  At the ILC, these results may be improved by an
order of magnitude in some cases by tuning the beam energy to produce
slow sleptons.  Precision studies of slepton decays are therefore
possible and will provide direct observations of gravitational effects
at colliders; percent level measurements of the gravitino mass and
Newton's constant; precise determinations of the gravitino's
contribution to dark matter and supersymmetry breaking's contribution
to dark energy; quantitative tests of supergravity relations; and
laboratory studies of Big Bang nucleosynthesis and cosmic microwave
background phenomena.
\end{abstract}

%04.65.+e Supergravity
%12.60.Jv Supersymmetric models
%26.35.+c Big Bang nucleosynthesis
%98.80.Es Observational cosmology (including Hubble constant, 
%          distance scale, cosmological constant, early Universe, etc)

\maketitle

\section{Introduction}
\label{sec:introduction}

Weak-scale supersymmetry remains a beautiful framework for resolving
the problems of electroweak symmetry breaking and dark
matter~\cite{Feng:2003zu}, and its discovery is among the most eagerly
anticipated events in particle physics.  Opportunities for
supersymmetry discoveries and studies at colliders depend largely on
which superpartner is the lightest supersymmetric particle (LSP).  In
simple supergravity models, supersymmetry is transmitted to standard
model superpartners through gravitational interactions, and
supersymmetry is broken at a high scale.  The mass of the gravitino
$\gravitino$ is
\begin{equation}
m_{\gravitino} = \frac{F}{\sqrt{3} \mstar} \ ,
\label{gravitinomass}
\end{equation}
and the masses of standard model superpartners are
\begin{equation}
\tilde{m} \sim \frac{F}{\mstar} \ ,
\label{tildem}
\end{equation}
where $F \sim (10^{11}~\gev)^2$ is the supersymmetry breaking scale
squared, and $\mstar = (8 \pi G_N)^{-1/2} \simeq 2.4 \times
10^{18}~\gev$ is the reduced Planck scale.  The precise ordering of
masses depends on unknown, presumably ${\cal O}(1)$, constants in
\eqref{tildem}.  Most supergravity studies assume that the LSP is a
standard model superpartner, such as a slepton or neutralino.

Recently attention has turned to the other logical possibility,
namely, that the gravitino is the LSP~\cite{Feng:2003xh,Feng:2003uy,%
Feng:2003nr,Ellis:2003dn,Buchmuller:2004rq,Feng:2004zu,Feng:2004mt,%
Wang:2004ib,Feng:2004gn,Ellis:2004bx,Roszkowski:2004jd}.  In
supergravity where supersymmetry breaking is mediated by gravity, the
gravitino has a mass $m_{\gravitino} \sim 100~\gev$ if the
superpartner mass scale is $\tilde{m} \sim 100~\gev$. The gravitino
couplings are also suppressed by $\mstar$.  The gravitino's extremely
weak interactions imply that it is irrelevant for most supersymmetric
processes.  However, if the gravitino is the LSP, the next-to-lightest
supersymmetric particle (NLSP) decays to its standard model partner
and a gravitino.  The NLSP is a weak-scale particle decaying
gravitationally and so has a natural lifetime of
\begin{equation}
\frac{\mstar^2}{\tilde{m}^3} \sim 10^4 - 10^8~\s \ ,
\label{lifetime}
\end{equation}
as will be discussed more fully in \secref{slepton}.  This lifetime
emerges naturally in this simple supersymmetric scenario.  At the same
time, it is outlandishly long by particle physics
standards~\cite{Covi:1999ty}.  It requires a revamping of many aspects
of supersymmetric phenomenology and cosmology and opens up novel
opportunities, including the one discussed here.

The gravitino LSP scenario is constrained by cosmological and
astrophysical data.  The gravitino is a stable superweakly massive
particle (superWIMP) and forms dark matter. Its production during
reheating and by NLSP decays is therefore constrained by measurements
of the non-baryonic cold dark matter density.  NLSP decays also
deposit electromagnetic~\cite{Kawasaki:1994sc,Holtmann:1998gd,%
Kawasaki:2000qr,Cyburt:2002uv} and
hadronic~\cite{Jedamzik:2004er,Kawasaki:2004yh} energy into the
universe well after Big Bang nucleosynthesis (BBN), and so may ruin
the successful predictions of standard BBN.  These decays may also
distort the cosmic microwave background (CMB) from its observed
Planckian spectrum.  Last, photons produced in NLSP decays are subject
to bounds on the diffuse photon flux.

The impact of these constraints on the gravitino LSP scenario have
been considered in detail.  In addition to the leading two-body NLSP
decays to the
gravitino~\cite{Feng:2003xh,Feng:2003uy,Feng:2003nr,Ellis:2003dn},
three-body NLSP decays must also be considered when they are the
leading contribution to hadronic
cascades~\cite{Feng:2004zu,Feng:2004mt}.  The result is that the
gravitino LSP scenario is not excluded and, in fact, all constraints
may be satisfied for natural weak-scale NLSP and gravitino masses.

Not all possibilities are allowed, however, and two results are
particularly worth noting.  First, neutralino NLSPs are highly
disfavored~\cite{Feng:2004mt,Roszkowski:2004jd}.  Neutralinos
typically have two-body decays $\chi \to Z \gravitino \to q \bar{q}
\gravitino$.  The resulting hadronic cascades destroy BBN successes,
and exclude this scenario unless such decays are highly suppressed.
Kinematic suppression is not viable, however --- if $m_{\chi} -
m_{\gravitino} < m_Z$, the decay $\chi \to \gamma \gravitino$ takes
place so late that it violates bounds on electromagnetic cascades.
Neutralino NLSPs are therefore allowed only when the two-body decays
to $Z$ bosons are suppressed dynamically, as when the neutralino is
photino-like, a possibility that is not well-motivated by high energy
frameworks.  Slepton and sneutrino NLSPs also produce hadronic energy
when they decay, but this occurs only through three-body decays.
These have been analyzed and found to be
safe~\cite{Feng:2004zu,Feng:2004mt}.  As a result, the most natural
NLSP candidates are sleptons, particularly the right-handed stau.

Second, cosmological constraints exclude the upper range of lifetimes
in \eqref{lifetime}~\cite{Feng:2003xh,Feng:2003uy}.  Very late decays
occur in a cold universe where decay products are not effectively
thermalized and so are especially dangerous.  For typical thermal
relic NLSP abundances, the CMB and BBN constraints therefore provide
an upper bound on NLSP lifetimes, roughly excluding those above a
year.

In passing, we note that the scenario outlined above has a number of
other motivations.  One such motivation is from BBN.  Late NLSP decays
not only pass all BBN constraints, they may even resolve the leading
BBN anomaly by destroying $^7$Li to bring the predicted abundance in
line with the low values favored by
observations~\cite{Feng:2003xh,Feng:2003uy}.  To resolve the $^7$Li
anomaly, the preferred NLSP lifetime is $\sim 3 \times
10^6~\s$~\cite{Ellis:2003dn}, that is, about a month.  A second
motivation follows from considerations of
leptogenesis~\cite{Fukugita:1986hr}. Gravitinos may be produced during
reheating.  If the gravitino is not the LSP, its late decays are
dangerous to BBN, and require reheating temperatures $T_{\text{RH}}
\alt 10^5~\gev$ to $10^8~\gev$~\cite{Kawasaki:2004yh}, in conflict
with the requirement $T_{\text{RH}} \agt 3 \times 10^9~\gev$ of
thermal leptogenesis~\cite{Buchmuller:2003gz}.  In contrast, in the
gravitino LSP scenario, the gravitino does not decay, and the reheat
temperature is bounded only by the overclosure constraint on the
gravitino density.  For $m_{\gravitino} \sim 100~\gev$, reheat
temperatures as high as $\sim 10^{10}~\gev$ are
allowed~\cite{Moroi:1993mb,Bolz:2000fu}, consistent with thermal
leptogenesis~\cite{Bolz:1998ek,Fujii:2003nr}.  Additional connections
between leptogenesis and gravitino LSPs are discussed in
Ref.~\cite{Allahverdi:2004ds}.

Given all of these motivations, we investigate here the collider
implications of a gravitino LSP with a charged slepton NLSP with
lifetime under (but not much under) a year~\cite{Goity:1993ih}.  In
particular, we investigate the possibility of trapping sleptons in
material placed just outside Large Hadron Collider (LHC) or
International Linear Collider (ILC) detectors.  This material may then
be moved to some quiet location so that slepton decays may be observed
in a relatively background-free environment.  Although these
objectives may be realized in many ways, we study here the
particularly simple possibility of trapping sleptons in water tanks
which may be drained periodically to underground reservoirs where the
slepton decays may be observed.

In \secref{slepton} we discuss the relevant properties of sleptons in
the gravitino LSP scenario.  In \secref{trap} we discuss our procedure
for maximizing the number of sleptons trapped given a fixed volume of
water.  This is applied to the cases of the LHC and ILC in
\secsref{LHC}{ILC}, respectively.  At the LHC, we find that tens to
thousands of sleptons may be trapped each year, depending on the
overall mass scale of supersymmetry.  At the ILC, typically far fewer
sleptons are produced.  However, by controlling the beam energy, the
velocity of produced sleptons may be tuned to some low value, allowing
a large fraction of sleptons to be trapped.  By exploiting this
feature, we find that an order of magnitude more sleptons may be
trapped at the ILC than at the LHC.  These results imply that percent
level studies of slepton decays may be possible.  Such studies will
have fundamental implications for supergravity, supersymmetry
breaking, dark matter, and dark energy.  These implications and our
conclusions are discussed in \secref{implications}.

\section{Slepton Properties in the Gravitino LSP Scenario}
\label{sec:slepton}

\subsection{Slepton Mass}

In supergravity with a gravitino LSP, the slepton NLSP is expected to
have a weak-scale mass.  Current collider bounds require
$m_{\tilde{l}} > 99~\gev$ from null searches for long-lived charged
tracks at LEP II~\cite{leplimit}.

Cosmology brings additional considerations, however.  Gravitinos
produced in the late decays of sleptons are superWIMP dark matter.
Barring the possibility of entropy production after slepton freeze
out, the gravitino relic density must therefore satisfy
\begin{equation}
\Omega_{\gravitino} h^2 = \frac{m_{\gravitino}}{m_{\slepton}} 
\Omega_{\slepton}^{\text{th}} h^2 < \Omega_{\text{DM}} h^2 \ ,
\end{equation}
where $\Omega_{\slepton}^{\text{th}}$ is the slepton's thermal relic
density, and the non-baryonic cold dark matter density is constrained
to the range $0.094 < \Omega_{\text{DM}} h^2 < 0.124$.  Assuming
$m_{\gravitino}$ and $m_{\slepton}$ are not too disparate, this
provides an upper bound on the slepton and gravitino masses, since
$\Omega_{\slepton}^{\text{th}} \propto m_{\slepton}^2$.  Without
special effects, $\Omega_{\slepton}^{\text{th}} \sim
\Omega_{\text{DM}}$ for $m_{\slepton} \sim 700-1000~\gev$.  If
$\frac{m_{\gravitino}}{m_{\slepton}} \agt 0.1$, the overclosure
constraint requires superpartner masses below about 3 TeV.

On the other hand, a particularly attractive possibility is that
gravitino superWIMPs are most or even all of the non-baryonic dark
matter.  Although not a strict requirement, one might therefore prefer
$\Omega_{\gravitino} \approx \Omega_{\text{DM}}$.  The decay to the
gravitino only reduces the relic density.  Without special effects,
then, overclosure requires $m_{\slepton} \agt 700-1000~\gev$.  Such
heavy sleptons will be difficult to explore at the LHC and are
kinematically inaccessible in the first stage of the ILC.

Just as in conventional neutralino dark matter scenarios, however,
there are supplementary mechanisms for gravitino production.  One such
mechanism is co-annihilation.  If a neutralino $\chi$ is just slightly
heavier than the slepton, it will freeze out with the slepton and
later decay to the slepton, adding its thermal relic density to the
slepton's.  This allows supersymmetric models with much lower slepton
masses to produce the correct gravitino superWIMP dark matter density.
For example, in minimal supergravity with $A_0 = 0$, $\tan \beta =
10$, and $\mu >0$, the desired relic density may be achieved near the
$\stau$ LSP--$\chi$ LSP border at $\mgaugino = 300~\gev$, where
$m_{\stau} \approx m_{\chi} \approx 120~\gev$~\cite{Belanger:2004ag}.
Alternatively, gravitinos may be produced during reheating.  For
reheating temperatures $T_{\text{RH}} \sim 10^9~\gev$, as might be
preferred for leptogenesis as discussed above, gravitinos may again be
all of the non-baryonic dark matter for slepton masses as low as 120
GeV~\cite{Roszkowski:2004jd}.  It is clear that such effects may be
very important.  We will consider a variety of slepton masses below,
including those within reach of a first stage ILC.  We reiterate that
all of these considerations depend on the assumption that gravitinos
make up all of the dark matter. {}From a purely particle physics
viewpoint, this is clearly optional --- some other particle, such as
the axion, may be the dark matter.  In this case, gravitino LSPs and
light NLSP sleptons are perfectly possible without any additional
restrictions.

\subsection{Slepton Lifetime}

The width for the decay of a slepton to a gravitino is
\begin{equation}
 \Gamma(\tilde{l} \to l \tilde{G}) =\frac{1}{48\pi \mstar^2}
 \frac{m_{\tilde{l}}^5}{m_{\tilde{G}}^2} 
 \left[1 -\frac{m_{\tilde{G}}^2}{m_{\tilde{l}}^2} \right]^4 \ ,
\label{sfermionwidth}
\end{equation}
assuming the lepton mass is negligible. This decay width depends on
only the slepton mass, the gravitino mass, and the Planck mass.  In
many supersymmetric decays, dynamics brings a dependence on many
supersymmetry parameters.  In contrast, as decays to the gravitino are
gravitational, dynamics is determined by masses, and so no additional
parameters enter.  In particular, there is no dependence on left-right
mixing or flavor mixing in the slepton sector.

\subsection{Slepton Range in Matter}

Last, it will be crucial to this study to know the range of sleptons
in matter.  Charged particles passing through matter lose energy by
emitting radiation and by ionizing atoms.  At lower energies,
ionization dominates the energy loss, while at high energies,
radiation is the dominant effect.  As we will see below, it is
unreasonable to expect to stop sleptons with momenta much larger than
their rest mass.  For the present case, then, ionization losses are
dominant, and radiation is negligible.

The average energy loss due to ionization is given by the Bethe-Bloch
equation.  The low energy approximation to the Bethe-Bloch equation
may be derived~\cite{Ziegler:1999} by first calculating the
classical cross section for a collision with fixed impact parameter
and energy loss.  One then integrates the impact parameter from the
Compton wavelength of the free electron as seen by the charged
particle to a maximal impact parameter where the particle cannot
``see'' the electron in the time that it passes by the atom.  A more
refined treatment yields the Bethe-Bloch
equation~\cite{Bichsel:2002cf}
\begin{equation}
\frac{dE}{dx}
= K z^2 \frac{Z}{A} \frac{1}{\beta^2}
\left[ \ln \left( \frac{2 m_e c^2 \beta^2 \gamma^2}
{I \sqrt{1 + \frac{2m_e\gamma}{M} + \frac{m_e^2}{M^2}}}\right) -
\beta^2 - \frac{\delta}{2} \right] \ ,
\label{BB}
\end{equation}
where $dE/dx$ is the energy loss per $\g~\cm^{-2}$, $K =
0.307075~\mev~\g^{-1}~\cm^2$, and $m_e$ is the electron mass.  The
material is characterized by its atomic charge $Z$ in units of $e$;
its average nucleon number $A$; and its mean ionization energy $I$,
which is given for various elements in Ref.~\cite{Bichsel:2002cf}.
The incoming particle has mass $M$, charge $z$ in units of $e$,
velocity $\beta$, and dilation factor $\gamma = (1 - \beta^2)^{-1/2}$.

The parameter $\delta$ accounts for the fact that incoming particles
polarize the surrounding medium.  At high energies, this effect may be
included by setting
\begin{equation}
\delta = \Theta(E-E_0) \left[ \ln\left(\frac{E^2}{M^2}-1\right)
  + \ln\left(\frac{\hbar\omega_p}{I}\right)
  - \frac{1}{2} \right] \ ,
\label{density}
\end{equation}
where $\omega_p$ is the plasma frequency, and $E_0$ is the energy at
which the effect of polarization is significant.  This correction is
typically significant for $\beta \gamma \agt 10$.  We include this
effect, although, as we will see, sleptons that may be trapped in a
reasonably sized detector have $\beta \gamma < 1$, and so this effect
is also irrelevant for the final results of this study.

At low momentum, corrections to \eqref{BB} arise from the fact that
electrons in matter are bound to atoms.  This implies that there is
transverse momentum in the collision (Bloch corrections) and that the
electron may have momentum comparable to the incident particle (shell
corrections).  These corrections are significant only when the
momentum of the incoming particle is comparable to the electron
momentum.  As discussed below, we will neglect contributions to the
slepton range from very low velocities.  For ${\cal O}(100)~\gev$
sleptons traveling at the velocities we include, the Bloch and shell
corrections may be safely neglected.  The Barkas effect, which
introduces a dependence on the sign of the charge of the incoming
particle is also significant only for very low $\beta$, and may be
safely neglected for the velocities we include.

The Bethe-Bloch equation \eqref{BB} with $\delta$ as given in
\eqref{density} is accurate down to $\beta \sim 0.05$. Below this
velocity, experimental data are fit by parameterization schemes.  We
do not have this luxury because, of course, there are as yet no
experimental data for sleptons.  Although we could use the
parameterizations adopted for standard model particles, it has been
found that the heavier the particle the worse these models are at
describing the low momentum behavior~\cite{Tai:1997}.

Rather than grapple with how sleptons should behave at $\beta < 0.05$,
we adopt the following procedure for determining the slepton range in
matter.  For a certain low velocity, $dE/dx$ in \eqref{BB} peaks and
then rapidly drops to zero.  We denote this velocity
$\beta_{\text{peak}}$; its value is typically $\sim 0.01$. At
$\beta_{\text{peak}}$, the value of $dE/dx$ is very high, and we
expect that the distance the slepton travels as it slows from
$\beta_{\text{peak}}$ to thermal equilibrium is negligible relative to
the distance it traveled in slowing down from its initial velocity to
$\beta_{\text{peak}}$.  We further take the continuous slowing down
approximation.  With these assumptions, the range $R$ in $\g~\cm^{-2}$
for a slepton with energy $E'$ is
\begin{equation}
  R( E' ) = \frac{1}{K} \frac{A}{Z}
  \int_{M + \delta M}^{E'} dE
  \frac{ \frac{M^2}{E^2}-1} {\ln\left(\frac{2m_ec^2}{I} 
\frac{E^2-M^2}{\sqrt{M^2+2 E m_e+m_e^2}}\right)+
\left(\frac{M^2}{E^2}-1\right) - \delta} \ ,
\end{equation}
where $\delta$ is given in \eqref{density}, and $M + \delta M$ is the
energy at which the slepton has velocity $\beta_{\text{peak}}$.  The
values used for $\delta M$, $I$, $E_0$, and $\hbar\omega_p$ for the
particular case of a 219 GeV slepton are given in
\tableref{parameters}.

\begin{table}[tb]
\caption{Range parameters for lead and water, assuming a 219 GeV
slepton.
\label{table:parameters}
}
\begin{tabular}{|c|c|c|c|c|c|}
\hline Material 
& \quad $\delta M \ (\mev)$ \quad 
& \quad $I \ (\ev)$ \quad
& \quad $E_0 \ (\tev)$ \quad 
& \quad $\hbar \omega_p \ (\ev)$ \quad \\ \hline
Lead  & 110 & 820 & 4.4 & 61 \\ \hline
Water & 220 &  75 & 1.3 & 21 \\ \hline 
\end{tabular}
\end{table}

The Bethe-Bloch equation gives an average value of the stopping power,
but there are always fluctuations about these values.  For thin
materials the most probable energy loss may be very different from the
mean energy loss~\cite{Bichsel:2002cf}.  However, if the path length
in material is large enough, the energy loss distribution is Gaussian.
The criterion for Gaussianity has been found in
Ref.~\cite{Schorr:1973} to be
\begin{equation}
\kappa = \frac{ \frac{K}{2}
\frac{Z}{A} \frac{1}{\beta^2}
\rho x}{T} > 10 \ ,
\end{equation}
where
\begin{equation}
T =
\frac{2m_ec^2\beta^2\gamma^2} 
{\sqrt{1+\frac{2m_e\gamma}{M}+\frac{m_e^2}{M^2}}} \ ,
\end{equation}
and $x$ is the path length.  For LHC and ILC detectors, $\kappa \agt
{\cal O}(100)$.  Given that the energy distribution of produced
sleptons is not wildly fluctuating, inclusion of this Gaussian range
distribution will have a negligible impact on our results, and we do
not include it.

\begin{figure}
\resizebox{6.5in}{!}{
\includegraphics{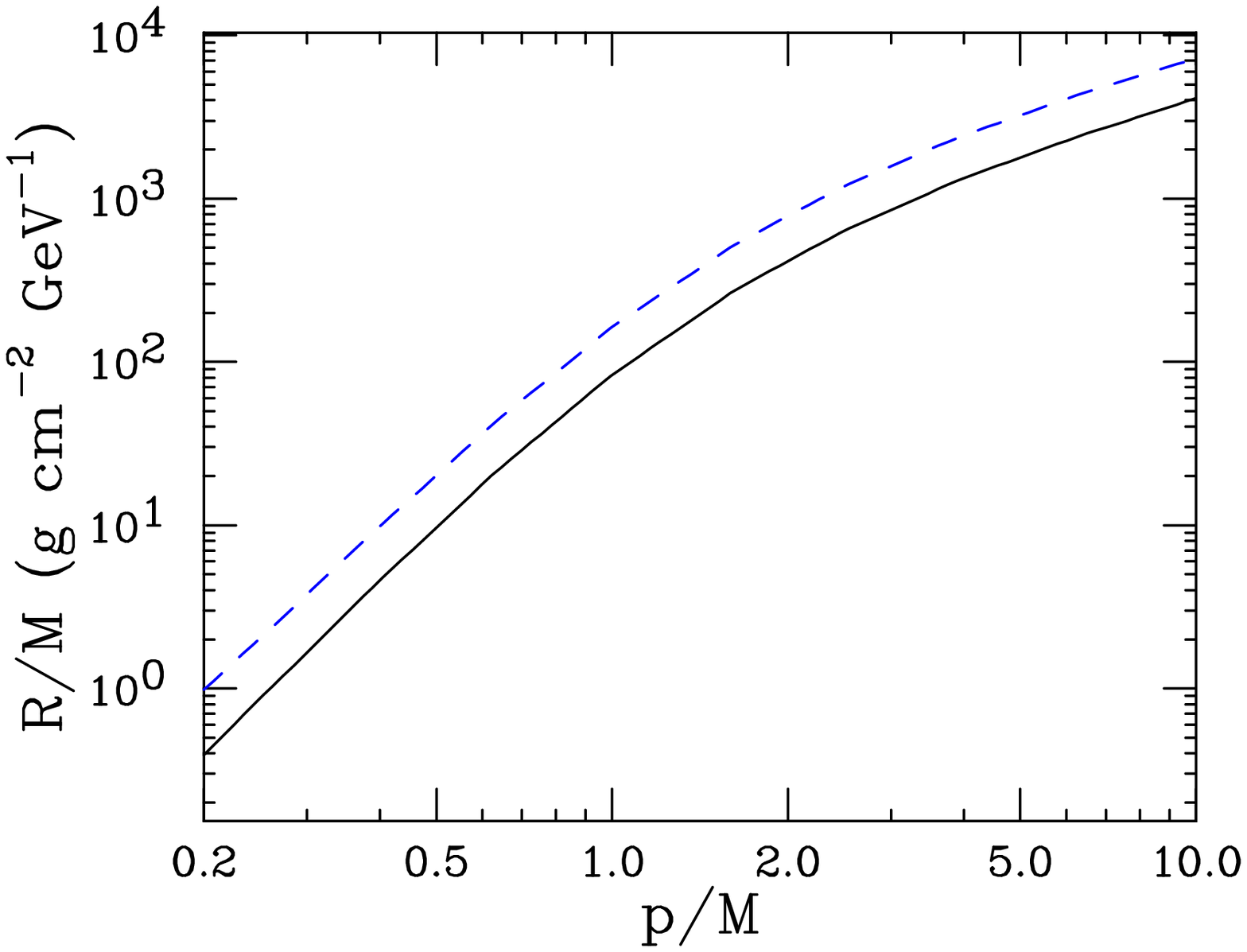} \qquad
\includegraphics{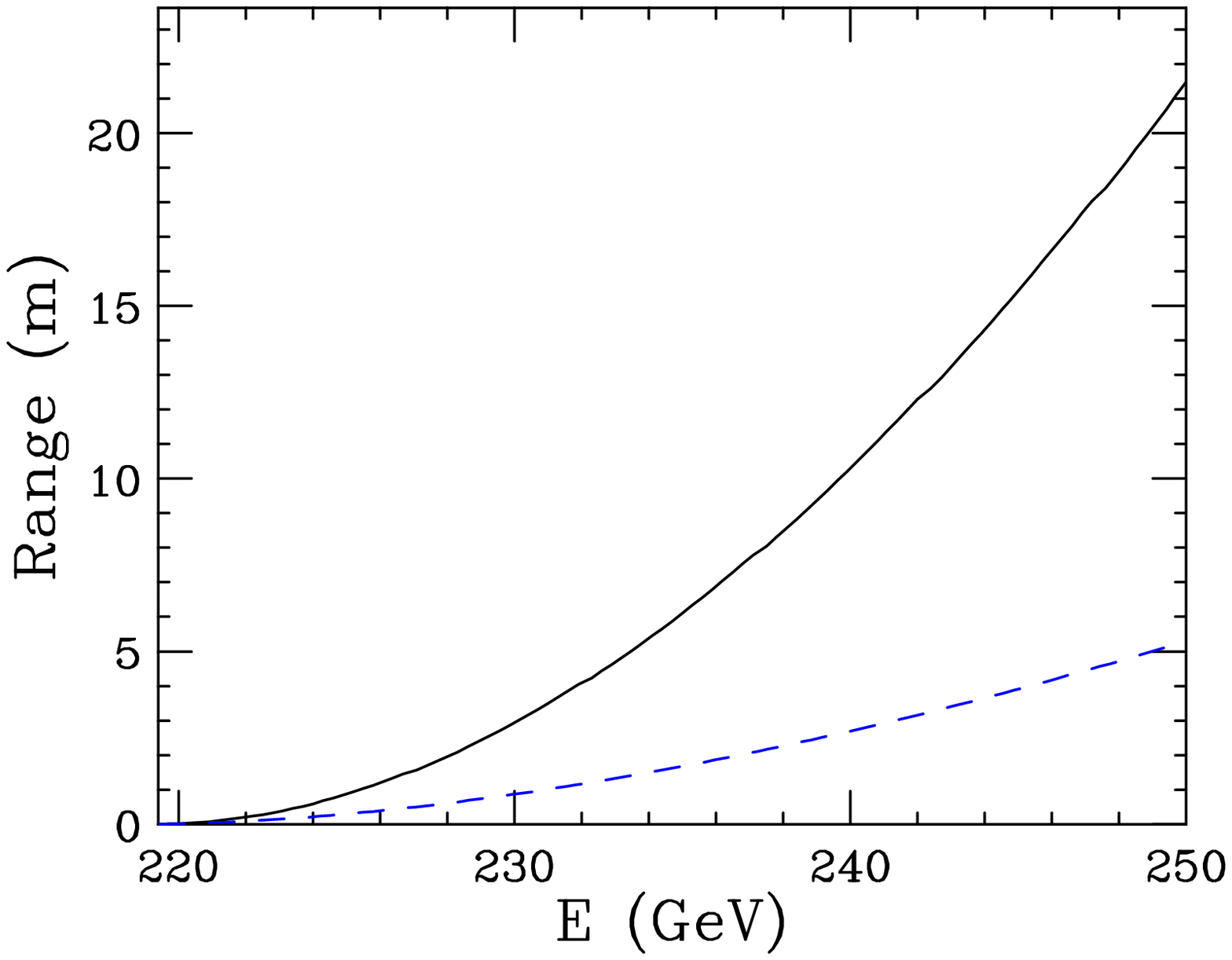}
}
\caption{The range to mass ratio $R/M$ as a function of $p/M = \beta
\gamma$ (left), and, for the specific case of a slepton with mass
219 GeV, the range as a function of energy (right).  Results are
given for water (solid) and lead (dashed).
\label{fig:range} }
\end{figure}

Using the procedure described above, the range to mass ratios $R/M$
for lead and water are given in \figref{range}.  These results agree
beautifully with published results~\cite{Bichsel:2002cf}.  The range
as a function of energy for the particular case of a slepton with mass
219 GeV is also given. A slepton with this mass and energy 240 GeV
(250 GeV) travels about 10 meters (21 meters) in water before
stopping.

\section{Slepton Trap Optimization}
\label{sec:trap}

We will consider the possibility of trapping sleptons in water placed
just outside a collider detector.  Of course, any material may be
used, and our analysis, at the level we have pursued it, is valid for
any material.  We consider water to be promising, however, as it is
potentially feasible to increase the concentration of sleptons and/or
move it to a place where slepton decays may be observed in a
background-free environment.

We would like to optimize the placement and shape of the water tank.
If the gravitino LSP scenario is realized in nature, given the
implications described in \secref{implications}, we consider the
importance of slepton trapping studies to be sufficient to enlarge
detector halls.  We therefore do not consider constraints from
existing detector halls in considering trap geometries.  Even with
this simplification, however, a detailed discussion of optimization
requires careful accounting of the various LHC and ILC detector
component geometries, costs, and other factors.  

Rather than undertake such a detailed study, we consider here a simple
detector model to highlight the physics that may be explored in more
detail in following studies.  We characterize the properties of the
inner detector by two parameters: $\rin$, the distance from the
interaction point (IP) to the outside of the detector, and $\rinwe =
\int \! \rho \, dl$, the density-weighted distance between the IP and
the outside of the detector, typically measured in meters water
equivalent (mwe).  We will assume that $\rin$ and $\rinwe$ are
independent of polar angle $\theta$ and azimuthal angle $\phi$; that
is, we model the inner detector as spherically symmetric in both size
and material depth.

The amount of energy lost by a slepton traveling through a detector is
determined by $\rinwe$.  (For realistic detector sizes, the tracks of
sleptons with sufficient energy to pass through the detector have
negligible curvature.)  The radius of the LHC detectors is
approximately $12 \lambda_I$ in the direction perpendicular to the
beam line, where $\lambda_I$ is the nuclear interaction
length~\cite{ATLAS:ATLAS,CMS:CMS}.  This number depends on rapidity,
but we take this minimal value for our spherical detector.  The ILC
detector is expected to be slightly smaller, but sheets of lead or
other material can always be used to increase the effective radius to
that of the LHC detectors.  As we will see, this may be advantageous.
Given these considerations, we assume for this study that the energy
loss at the LHC and ILC is approximately that of a particle traveling
through $12 \lambda_I$ of water.  Since $\lambda_I = 83.6~\g~\cm^{-2}$
for liquid water, this sets $\rinwe = 10~\mwe$.

\begin{figure}
\resizebox{3.25 in}{!}{
\includegraphics{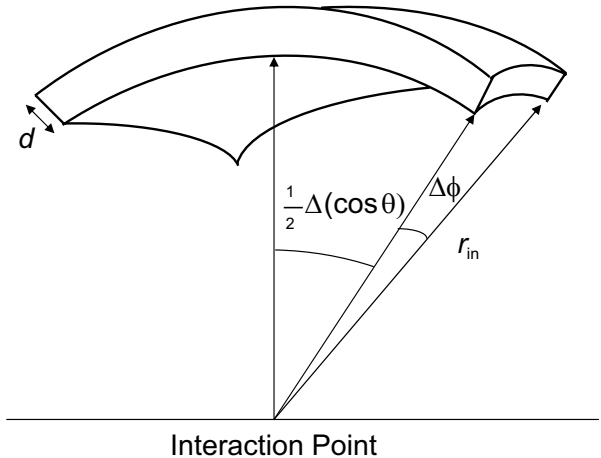}
}
\caption{Diagram of the slepton trap geometry.  The trap is assumed to
be a spherical shell with inner radius $\rin$, and depth $d$ as shown.
The angular parameters $\frac{1}{2} \Delta\left(\cos\theta\right)$ and
$\Delta\phi$ of \eqref{shape} are also indicated.
\label{fig:trapfig} }
\end{figure}

The number of trapped sleptons is, of course, maximized by placing the
water tank as close to the IP as possible.\footnote{This assumes that
no additional material is added between the detector and the trap.  It
may be advantageous to place the trap farther away if a material that
slows sleptons is added between the detector and the trap.  An example
is discussed in \secref{ILC}.}  In addition, as we will see, the polar
angle distribution of {\em slow} sleptons, that is, those that have a
chance of being trapped, is either uniform or peaked perpendicular to
the beam line at polar angle $\theta = \pi/ 2$.  To optimize the water
tank placement, then, we consider the family of tank geometries
specified by
\begin{eqnarray}
\rin < & r & < \rin + d \nonumber \\
& |\cos \theta | & < \frac{1}{2} \Delta (\cos \theta) 
\label{costheta} \nonumber \\
0 < & \phi & < \Delta \phi \ ,
\label{shape}
\end{eqnarray}
where $(r, \theta, \phi)$ are polar coordinates centered on the IP, 
and $d$ is the tank's radial depth.  This geometry is illustrated in
\figref{trapfig}.  All sleptons with range less than
$\rinwe$ are trapped in the detector.  However, all sleptons with
range between $\rinwe$ and $\rinwe + d$, and polar and azimuthal
angles in the ranges given in \eqref{shape} are caught in the water
tank.\footnote{In practice, for the Monte-Carlo simulations below, we
smooth distributions in $\phi$ by including all events that pass the
$r$ and $\cos\theta$ cuts with weight $\Delta \phi/2\pi$.}  The water
tank volume is
\begin{equation}
V = \frac{1}{3} \left[ \left( \rin + d \right)^3 -
  \rin^3 \right] \Delta (\cos \theta) \Delta \phi \ .
\label{volume}
\end{equation}

\begin{figure}
\resizebox{3.25 in}{!}{
\includegraphics{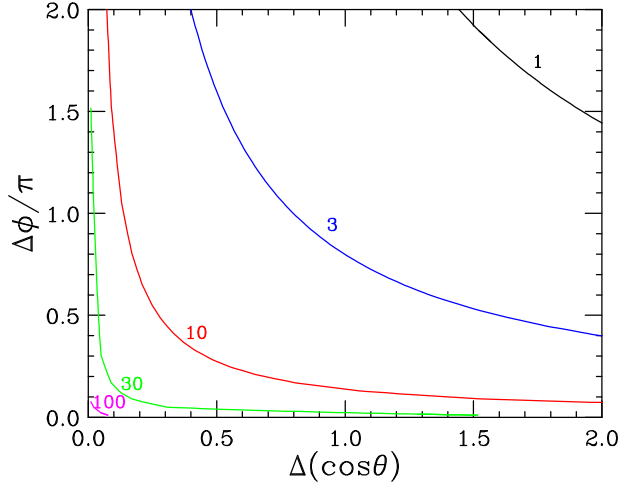}
}
\caption{The depth $d$ in meters of a 1 kton water trap in the
  $(\Delta (\cos \theta), \Delta \phi)$ plane for $\rin = 10~\m$.
\label{fig:depth} }
\end{figure}

In summary, the number of trapped sleptons is 
\begin{equation}
N(V, \Delta (\cos \theta), \Delta \phi, \rin, \rinwe) \ ,
\end{equation}
where the first four parameters determine the depth $d$ through
\eqref{volume}.  Sample depths for a 1 kton trap ($V =
1000~\mcubedwe$) are shown in \figref{depth}.  In the following
sections, we fix $\rin$ and $\rinwe$ to appropriate values and choose
three representative sizes $V$.  We then scan over all possible values
of $\Delta (\cos \theta)$ and $\Delta \phi$ to maximize $N$.  In this
way, we determine the optimal shape for the water tank and the maximal
number of sleptons that may be trapped.

\section{Slepton Trapping at the LHC}
\label{sec:LHC}

The LHC is scheduled to collide protons with protons at $\sqrt{s} =
14~\tev$ beginning in 2007-08.  Its initial luminosity is expected to
be $10~\ifb/\yr$, growing to $100~\ifb/\yr$.  We present results below
for $100~\ifb$, a high luminosity year.

The ATLAS and CMS detectors are cylindrical with radii 10 m and 8 m,
respectively.  As explained in \secref{trap}, we choose $\rin = 10~\m$
and $\rinwe = 10~\mwe$.  Although we consider only ATLAS and CMS,
other detectors could provide promising opportunities.  For example,
LHCb is an asymmetric detector and allows for smaller values of
$\rinwe$ and $\rin$.  This may make improved results possible and is a
possibility well worth considering.

In the gravitino LSP scenario, all supersymmetry events produce two
long-lived NLSP sleptons.  The dominant source of NLSPs at hadron
colliders is typically pair production of strongly interacting
superpartners.  The number of trapped sleptons is therefore
model-dependent in a complicated way, as it depends not only on the
slepton mass but also sensitively on the masses of colored
superpartners and their cascade decay patterns.

Here we consider minimal supergravity with the following parameters:
\begin{equation}
m_0 = 0 \ , \quad \mgaugino = 300-900~\gev \ , \quad A_0 = 0\ , \quad  
\tan\beta = 10 \ , \quad \mu>0 \ . 
\end{equation}
When the gravitino is not the LSP, this is in the excluded ``stau
LSP'' region. In the present scenario with a gravitino LSP, however,
these models are allowed, and this one-dimensional family of models
provides a simple set with which we can explore the prospects for
trapping sleptons at the LHC.  The lower bound on $\mgaugino$ is
determined by the requirement of a stau NLSP.  The number of trapped
staus rapidly diminishes as $\mgaugino$ increases, and the upper bound
on $\mgaugino$ is roughly where only a few staus may be trapped per
year.  The superpartner spectra for various $\mgaugino$ in these
models are given in \figref{spectrum}.

\begin{figure}
\resizebox{3.25 in}{!}{
\includegraphics{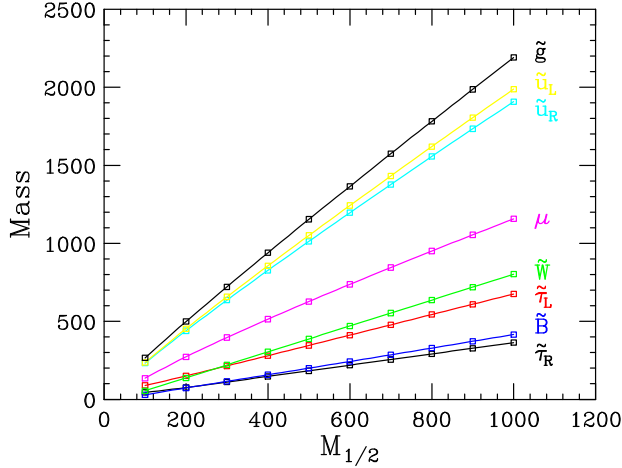}
}
\caption{Representative superpartner masses as a function of
  $\mgaugino$ in minimal supergravity with fixed $m_0 = 0$, $A_0 = 0$,
  $\tan\beta = 10$, and $\mu>0$.  The supersymmetry parameter $\mu$,
  which governs the Higgsino masses, is also shown.
\label{fig:spectrum} }
\end{figure}

The mass spectra of \figref{spectrum} are calculated by ISASUSY 7.69
with top mass $m_t = 175~\gev$~\cite{Paige:2003mg}.  We have also used
this package to generate LHC events. 100,000 non-standard model events
were generated for minimal supergravity models with the parameters
given above and $\mgaugino$ varying from 300 GeV to 900 GeV in 100 GeV
increments.  Helicity correlations are not included in ISAJET.
However, there are typically several steps in decay chains leading to
the NLSP, and so we do not expect helicity correlations to have a
significant impact on the NLSP distributions or on our final results.

In \figref{energy_600}, we show the energy distribution of NLSP staus
for the $\mgaugino = 600~\gev$.  Although many staus are produced,
most of these are extremely energetic and impossible to stop in a
reasonable distance.  Of course, a trap could be set up far from the
IP so that the intervening earth slows down the slepton, but such a
trap would be too far away to have a reasonable solid angle coverage
for any realistic volume $V$.

\begin{figure}
\resizebox{3.25 in}{!}{
\includegraphics{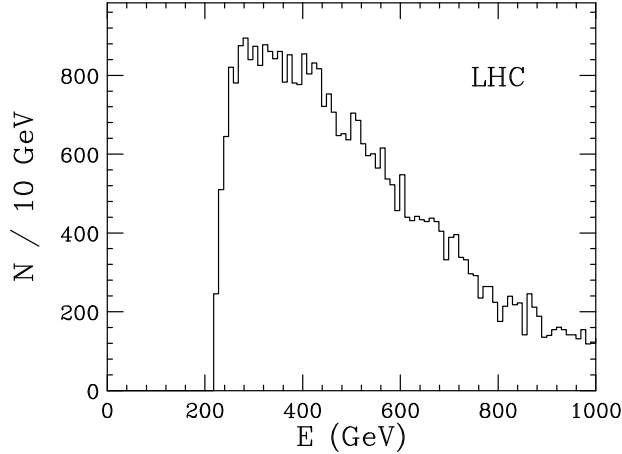}
}
\caption{The energy distribution of NLSP staus produced at the LHC for
  integrated luminosity $100~\ifb$ and minimal supergravity with $m_0
  = 0$, $\mgaugino = 600~\gev$, $A_0 = 0$, $\tan\beta = 10$, and
  $\mu>0$.  The NLSP stau mass is $219~\gev$.
\label{fig:energy_600} }
\end{figure}

The $\cos\theta$ distribution at $\mgaugino = 600~\gev$ is shown in
\figref{costheta_600}.  Given all produced NLSPs, the distribution is
strongly peaked in the beam directions.  However, imposing a cut on
energy, we find that the slow staus are produced roughly
isotropically.  Thus, for those staus that we might reasonably hope to
be trapped, any polar angle is as good as any other.  Given the
reality that LHC detectors are cylindrical, however, the closest a
trap may be placed to the IP is at $\cos\theta = 0$, justifying our
choice of centering our trap geometries at $\cos\theta = 0$, as
parameterized in \eqref{shape}.

\begin{figure}
\resizebox{3.25 in}{!}{
\includegraphics{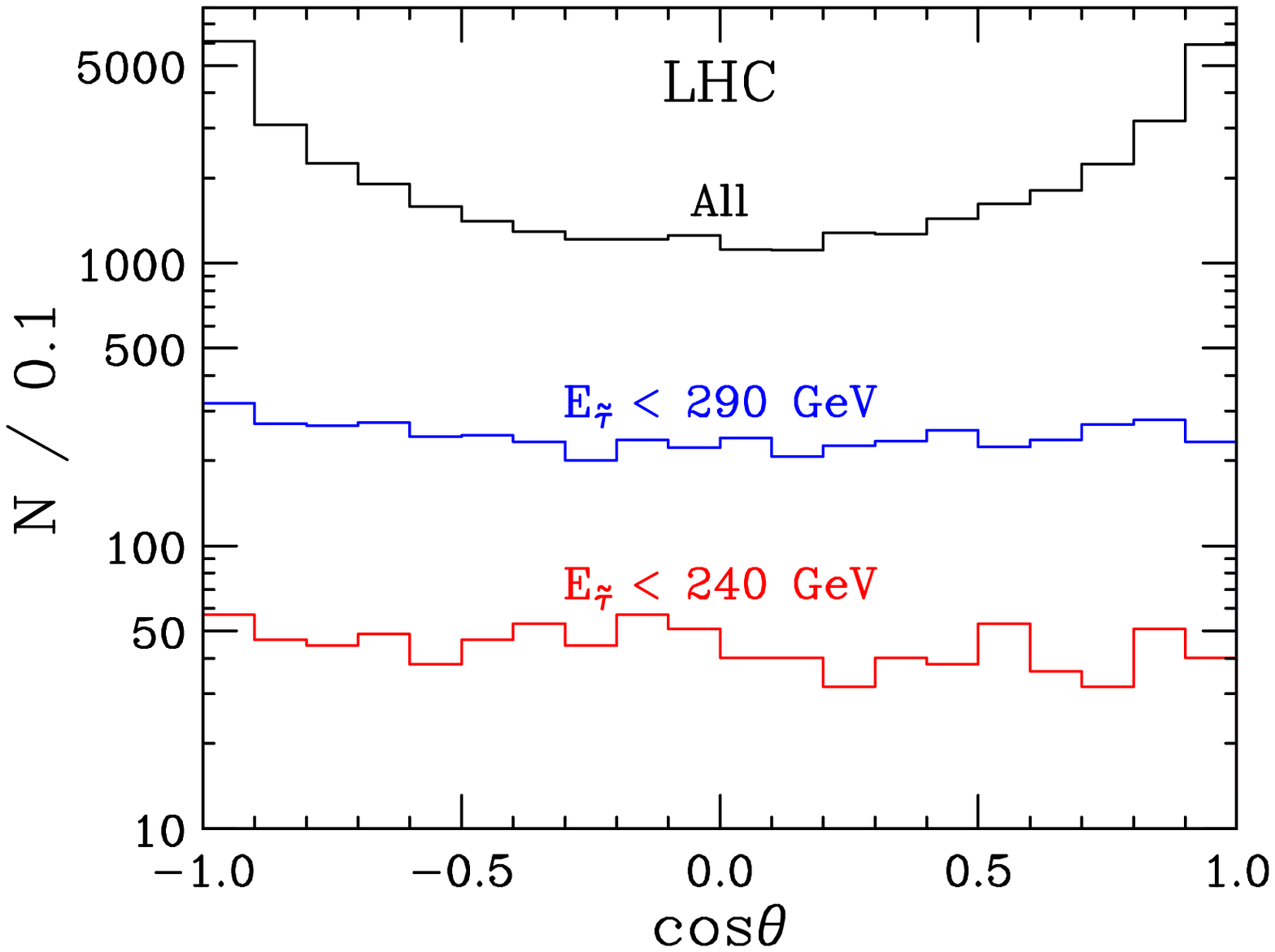}
}
\caption{The $\cos\theta$ distribution of NLSP staus produced at the
  LHC for integrated luminosity $100~\ifb$ and minimal supergravity
  with $m_0 = 0$, $\mgaugino = 600~\gev$, $A_0 = 0$, $\tan\beta = 10$,
  $\mu > 0$.  The NLSP stau mass is $219~\gev$. Distributions are
  given for all staus, $E_{\stau} < 290~\gev$ ($R < 100~\mwe$), and
  $E_{\stau} < 240~\gev$ ($R < 10~\mwe$). The total distribution is
  strongly peaked along the beam directions, but slow sleptons are
  produced isotropically.
\label{fig:costheta_600} }
\end{figure}

The number of trapped sleptons for optimized trap shape and placement
and various trap volumes is given in \figref{LHCresults}.  The trap is
optimized as described in \secref{trap}: we scan over all possible
$\Delta (\cos \theta)$ and $\Delta \phi$, and find the combination
that maximizes the number of sleptons that stop in the trap.

\begin{figure}
\resizebox{3.25 in}{!}{
\includegraphics{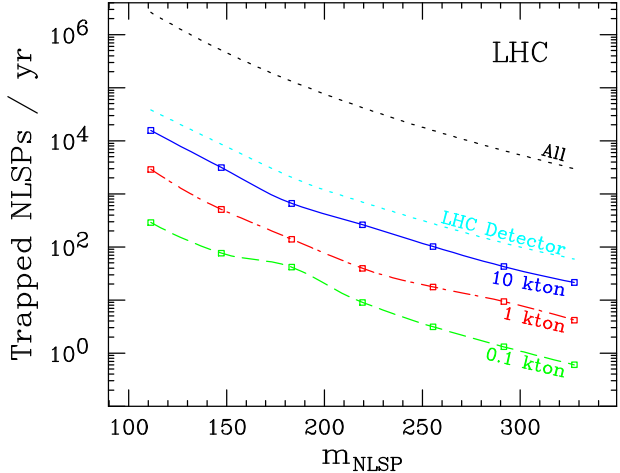}
}
\caption{The number of sleptons trapped per year at the LHC in water
tanks of size 10 kton (solid), 1 kton (dot-dashed), and 0.1 kton
(dashed).  The total number of sleptons produced is also shown (upper
dotted), along with the number of sleptons trapped in the LHC detector
(lower dotted).  The water tank shape and placement have been
optimized as described in the text.  These results assume luminosity
$100~\ifb/~\yr$, $\rin = 10~\m$, $\rinwe = 10~\mwe$, and minimal
supergravity models with $\mgaugino = 300, 400, \ldots, 900~\gev$,
$m_0 = 0$, $A_0 = 0$, $\tan\beta = 10$, and $\mu>0$.
\label{fig:LHCresults} }
\end{figure}

We find that only a small fraction of produced sleptons can be
trapped.  For example, for $\mgaugino = 600~\gev$, $4.2 \times 10^4$
NLSP sleptons are produced, but only 260, 40, and 9 are trapped in
water tanks of size 10, 1, and 0.1 kton, respectively.  For all models
considered, the 10 kton trap is optimized for $\Delta(\cos \theta) =
\Delta \phi / \pi = 2$; for this large volume, the trap is
sufficiently deep that there is little be gained by making the trap
deeper at the expense of solid angle coverage.  For $V= 1~\kton$, the
number of trapped sleptons is optimized by traps with
less-than-maximal angular coverage and depths of $d \approx 4~\m$.
Note that, because slow sleptons are produced isotropically, the
number of trapped sleptons depends, to a good approximation, on
$\Delta ( \cos \theta )$ and $\Delta \phi$ only through their product
$\Delta ( \cos \theta ) \Delta \phi$.

Despite the low efficiency for trapping, so many NLSPs are produced at
the LHC that significant numbers of NLSPs may still be trapped.  As
anticipated, the results are heavily dependent on the overall scale of
superpartner masses.  For $V = 10~\kton$ and $\mNLSP = 100 -
300~\gev$, the number of trapped sleptons varies from ${\cal O}(10^4)$
to ${\cal O}(10)$.  For the lighter sleptons considered, these results
imply that sufficient numbers of sleptons may be trapped to do
precision studies of slepton decay properties.  Assuming that slepton
decays may be observed in a background-free environment, we expect
percent level measurements of slepton decay widths.

Note that for all $\mgaugino$, a larger number of sleptons range out
in our spherically symmetric LHC detector than can be trapped in even
the 10 kton water trap.  For $\mgaugino = 600~\gev$, 700 are trapped
in the LHC detector itself.  The LHC detector has a large volume and
benefits from the fact that it begins at the IP and so has a large
angular coverage without sacrificing depth.  Unfortunately, it is not
clear to what extent these sleptons may be used --- their decays are
out of time, occur away from the IP, and take place in an environment
with significant cosmic ray background.  Given the large number of
sleptons that are automatically trapped in the LHC detector itself,
however, it is certainly worthwhile to explore ways to exploit them.

\section{Slepton Trapping at the ILC}
\label{sec:ILC}

In its first stage, the ILC will collide electrons and positrons with
center of mass energies up to 500 GeV.  In this first stage, the
luminosity has been estimated to be $340~\ifb/\yr$ for the TESLA
design~\cite{Brinkmann:2001qn} and $220~\ifb/\yr$ for the
NLC/JLC~\cite{Abe:2001nn}.  For this study, we assume luminosity
$300~\ifb/\yr$.  As in the LHC analysis above, we present results for
one year of running.

At present ILC detectors are expected to be slightly smaller than
their LHC counterparts. To be conservative, we assume $\rin = 10~\m$
and $\rinwe = 10~\mwe$, the same parameters we assumed in the LHC
case.  Of course, if the detector is smaller than this, it can always
be supplemented by adding plates of lead, for example, to mock up
these parameters.  As we will see, in the ILC case, such an approach
may in fact enhance our results very significantly.

At the ILC, scanning over supersymmetry models with a broad range of
superpartner mass scales, as done in the LHC analysis above, is not
particularly informative.  Models with heavy superpartners are simply
out of reach, and no sleptons may be produced, much less trapped.  On
the other hand, for models with superpartners within reach, the ILC
beam energy may be tuned to optimize the number of trapped sleptons,
to some extent offsetting variations in the scale of superpartner
masses in these models.  As we will see, the crucial feature is not
the exact mass of the slepton NLSP, but rather the presence of other
nearly degenerate superpartner states.

For the ILC, then, we limit our analysis to two models.  In the first,
which we denote ``NLSP only,'' the only superpartner within reach of
the ILC is an NLSP $\tilde{\tau}_R$ with mass 219 GeV.  This is
representative of the minimal case where the gravitino LSP scenario
may be probed at the ILC.  Of course, in many realistic models, there
are a number of other superpartners, notably other sleptons, fairly
degenerate with the NLSP.  We therefore consider also a second model,
which we denote ``mSUGRA,'' which is minimal supergravity with
$\mgaugino = 600~\gev$, $A_0 = 0$, $\tan\beta = 10$, and $\mu > 0$.
This model contains not only the 219 GeV $\tilde{\tau}_R$ of the
``NLSP only'' model, but also right-handed selectrons, right-handed
smuons, and a neutralino within the kinematic reach of a 500 GeV ILC.
The mSUGRA model is one of the family of models considered previously
in the LHC analysis, allowing us to compare the LHC and ILC at one
particular model point.  Because it contains the ``NLSP only'' model
as a subset, it also allows us to see the effect of having other
accessible and fairly degenerate superpartners.  The accessible
standard model superpartners of the two models and their masses are:
\begin{equation}
\left.
\begin{array}{l}
\hspace*{.44in} m_{\chi} \quad 242.9~\gev \\
m_{\tilde{e}_R}, m_{\tilde{\mu}_R} \quad 227.2~\gev \\
\hspace*{.36in} \left. m_{\tilde{\tau}_R} \quad 219.3~\gev 
\quad \right\} \ \text{NLSP only}  
\end{array} 
\quad \right\} \ \text{mSUGRA}
\end{equation}

We generate $10^4$ non-standard model ILC events for the mSUGRA model
with ISASUSY 7.69 with $m_t = 175~\gev$~\cite{Paige:2003mg}.  Events
for the ``NLSP only'' model are compiled by selecting the prompt stau
events from this event sample.  We choose beam width 0.12 mm, and
beamstrahlung parameter $\Upsilon = 0.1072$, and allow the subprocess
energy to vary over the entire range from $2 m_{\text{NLSP}}$ to
$\sqrt{s}$.

In the ``NLSP only'' model, the NLSP staus are produced through $e^+
e^- \to \gamma, Z \to \stau^+ \stau^-$. The stau energy distribution
is therefore given by the beam energy modified by initial state
radiation (ISR) and beamstrahlung. An example with $\sqrt{s} =
500~\gev$ is given in \figref{distributions_NLSP}.  The stau polar
angle distribution is also given in \figref{distributions_NLSP}.
Despite ISR and beamstrahlung, it retains the $\sin^2\theta$ shape of
the parton-level process.  The best place to trap sleptons is
therefore perpendicular to the beam line, justifying our choice of
centering our trap geometries at $\cos\theta = 0$, as parameterized in
\eqref{shape}.

\begin{figure}
\resizebox{6.5in}{!}{
\includegraphics{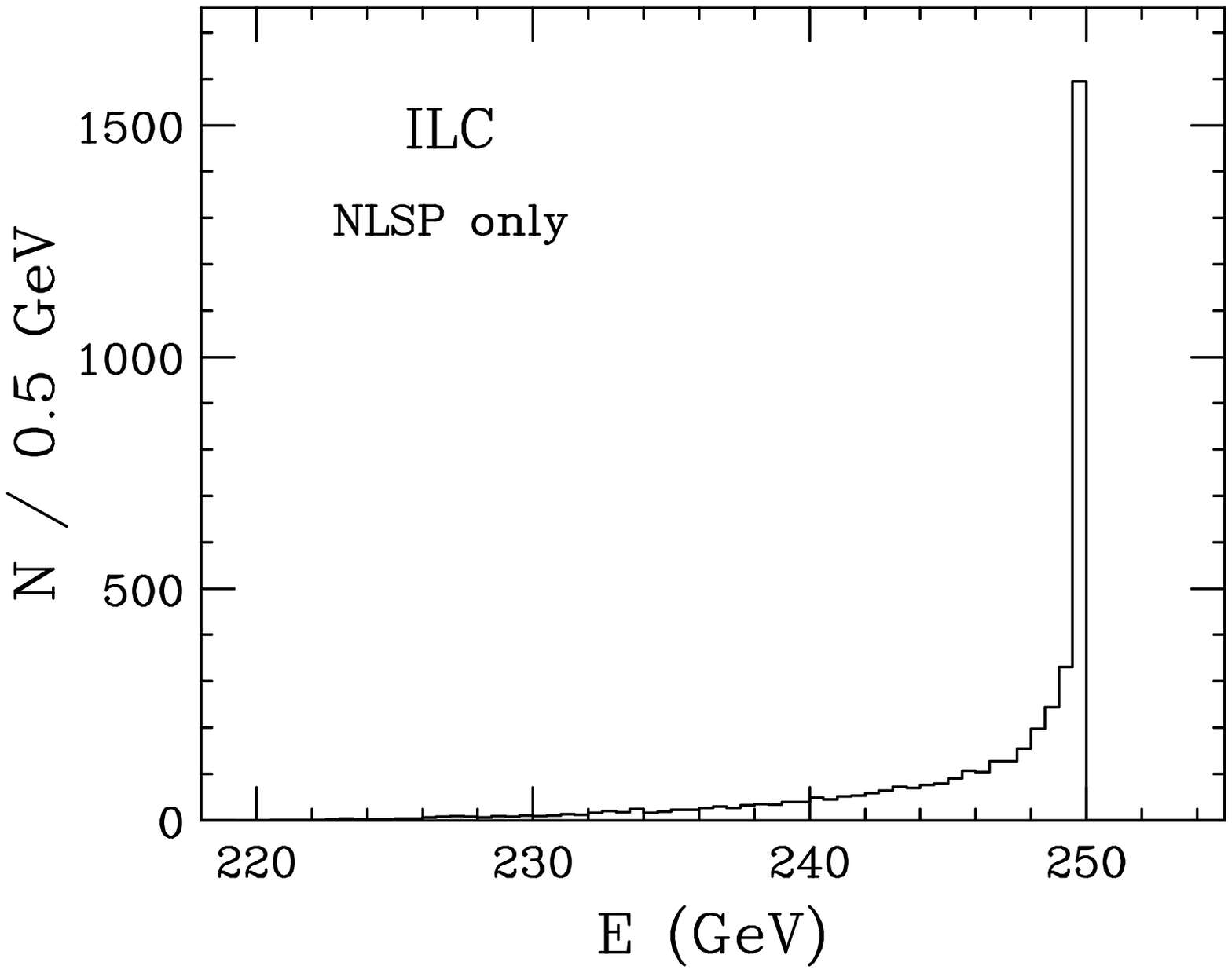} \qquad
\includegraphics{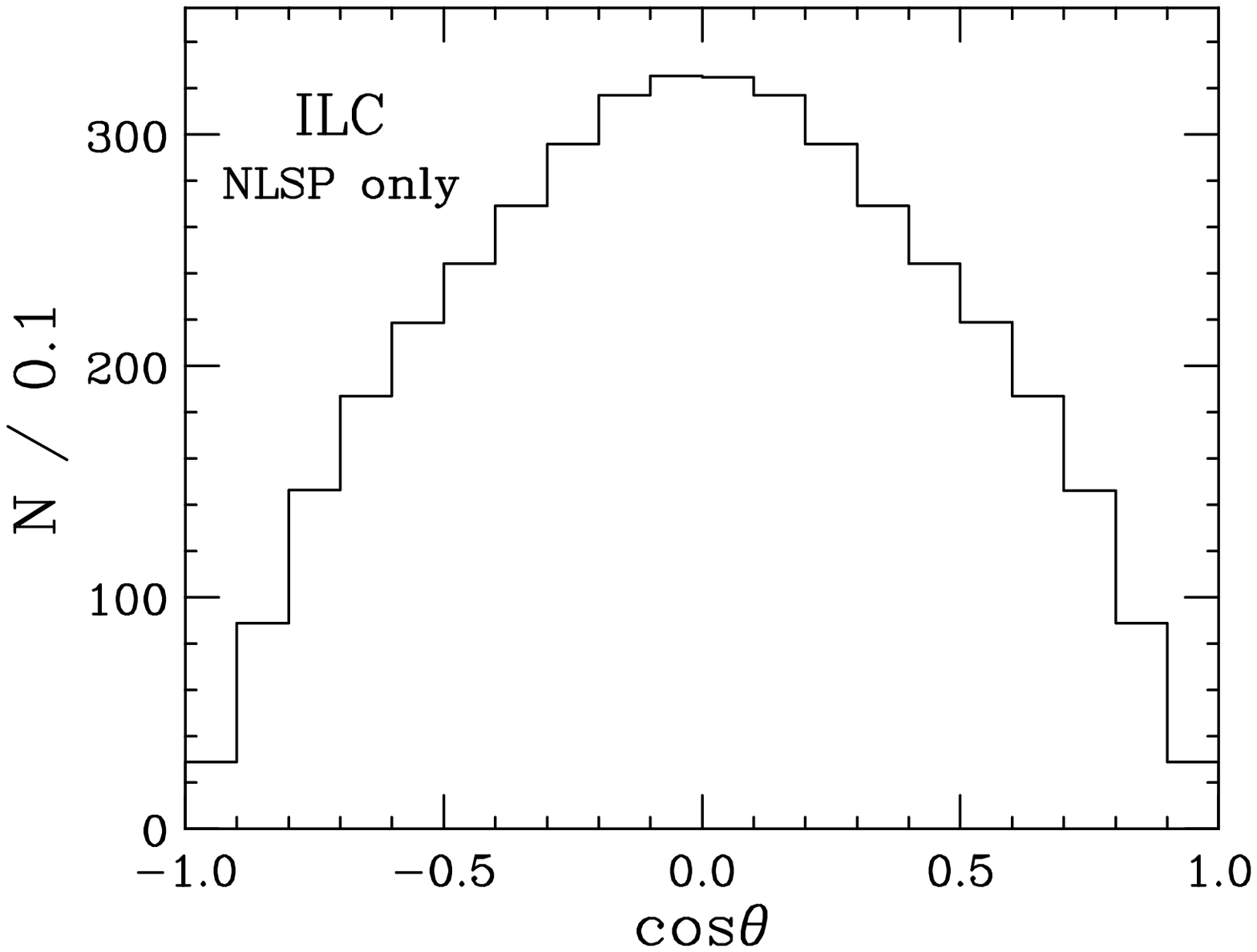}
}
\caption{Energy (left) and $\cos\theta$ (right) distributions for NLSP
  staus produced at the ILC assuming the ``NLSP only'' model, in which
  the only accessible superpartner is the NLSP stau with mass
  $219~\gev$.  Results are for $\sqrt{s} = 500~\gev$ and integrated
  luminosity $300~\ifb$.
\label{fig:distributions_NLSP} }
\end{figure}

The number of trapped sleptons for various trap sizes as a function of
center-of-mass energy $\sqrt{s}$ is given in \figref{ILCresults_NLSP}.
For $\sqrt{s} < 475~\gev$, no staus escape the ILC detector. At
$\sqrt{s} = 475~\gev$, however, sleptons in the sharp peak of the
energy distribution escape the ILC detector and may be caught in a
fairly thin water tank placed just outside the ILC detector.

\begin{figure}
\resizebox{3.25in}{!}{
\includegraphics{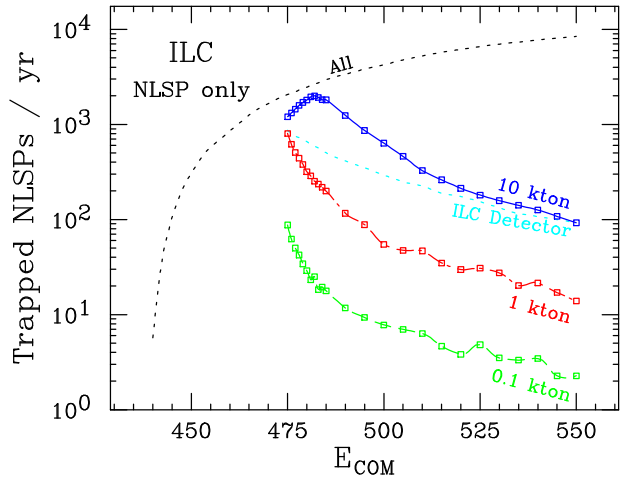} 
}
\caption{The number of sleptons trapped per year at the ILC in 10 kton
(solid), 1 kton (dot-dashed), and 0.1 kton (dashed) water traps.  The
total number of sleptons produced is also shown (upper dotted) along
with the number of sleptons trapped in the ILC detector (lower
dotted).  The trap shape and placement have been optimized, and we
assume $\rin = 10~\m$ and $\rinwe = 10~\mwe$, luminosity
$300~\ifb/\yr$ and the NLSP only model, where the only accessible
superpartner is a 219 GeV NLSP stau.
\label{fig:ILCresults_NLSP} }
\end{figure}

As evident in \figref{ILCresults_NLSP}, for the 1 and 0.1 kton water
traps, the number of trapped staus is maximized when the beam energy
is tuned to produce staus that just barely emerge from the ILC
detector.  The dependence on trap parameters is illustrated in
\figref{optimal_ILC_NLSP}.  The optimized trap configuration has
$\Delta (\cos\theta) \approx 1$ and $\Delta \phi = 2 \pi$; because the
stau distribution is peaked at $\cos\theta = 0$, it is beneficial to
sacrifice coverage at high rapidity to make the trap deeper.  For a 10
kton trap, the trap is sufficiently thick that the best results are
achieved for slightly higher beam energies where more of the
ISR/beamstrahlung tail may be caught.  For a 10 kton trap, we find
that the optimal trap configuration has $\Delta (\cos\theta) \approx
2$ and $\Delta \phi = 2 \pi$.

\begin{figure}
\resizebox{3.25in}{!}{
\includegraphics{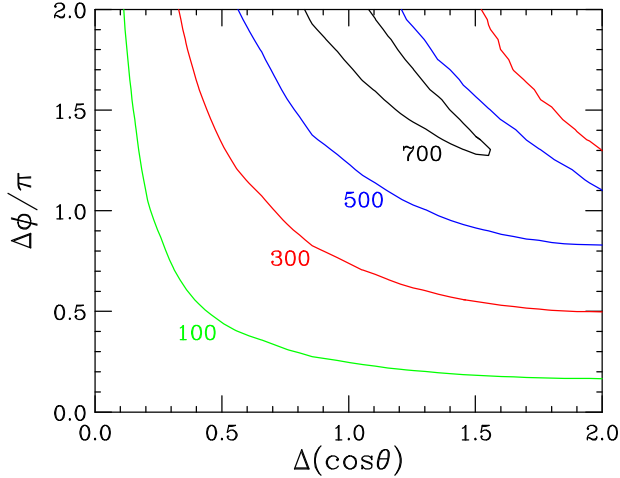}
}
\caption{The number of trapped staus at the ILC in the
  $(\Delta(\cos\theta), \Delta \phi)$ plane for the ``NLSP only''
  model, $\sqrt{s} = 475~\gev$, and integrated luminosity
  $300~\ifb/yr$.
\label{fig:optimal_ILC_NLSP} }
\end{figure}

For all trap sizes, however, the number of trapped sleptons is
maximized for beam energies near the 475 GeV threshold.  The tunable
beam energy and well-defined initial state are well-known virtues of
the ILC, but these are exploited in a qualitatively novel way here to
produce slow NLSP sleptons that may be easily caught.  Relative to the
case at the LHC, a much larger fraction of the produced staus can be
caught.  For example, of the 2650 staus produced at $\sqrt{s} =
482~\gev$ in a year, 2000 staus may be trapped in an optimized 10 kton
water trap!  Such results imply promising prospects for slepton
trapping even in the minimal case when the only superpartner
accessible at the ILC is the NLSP slepton.

Before considering the mSUGRA model, we note that the number of
produced staus continues to rise well beyond its value at $\sqrt{s} =
475~\gev$.  This suggests that our results may be improved
significantly by placing some dense material between the ILC detector
and the water tank.  By adding material depth to the ILC detector, the
threshold at which sleptons just barely emerge is moved to higher
$\sqrt{s}$ where the stau pair production cross section is higher.
For a dense material, such as lead, this can be achieved without
increasing $\rin$ much.  Such a strategy may in any case be required
to smooth out variations in $\rinwe$ inherent in realistic detectors.
Although we have not investigated this in detail, we expect that a
large enhancement may be possible.

We now turn to the mSUGRA model.  In this model, NLSP staus may again
be produced directly, but now they may also be produced in several
other ways: first, by $e^+ e^- \to \tilde{e}^+ \tilde{e}^-,
\tilde{\mu}^+ \tilde{\mu}^-$ followed by $\tilde{e} \to e \stau \tau$
and $\tilde{\mu} \to \mu \stau \tau$, and second, by $e^+ e^- \to \chi
\chi$, followed by $\chi \to \stau \tau$, or through the cascade $\chi
\to \tilde{e}, \tilde{\mu} \to \stau$.  The energy and $\cos\theta$
distributions of NLSP staus in the mSUGRA model are shown in
\figref{distributions_mSUGRA}.\footnote{Helicity correlations between
production and decay are not included in our event generation.  These
are, of course, absent for scalar particles, but may modify both the
energy and $\cos\theta$ distributions for staus produced through
$\chi$ pair production.}  As evident in \figref{distributions_mSUGRA},
the additional sources of staus increase the total number of staus
significantly, but just as significant, the cascade decays produce a
broad and flat tail in the energy distribution extending nearly down
to $m_{\stau}$.  The $\cos\theta$ distribution is nevertheless still
peaked at $\cos\theta = 0$.

\begin{figure}
\resizebox{6.5in}{!}{
\includegraphics{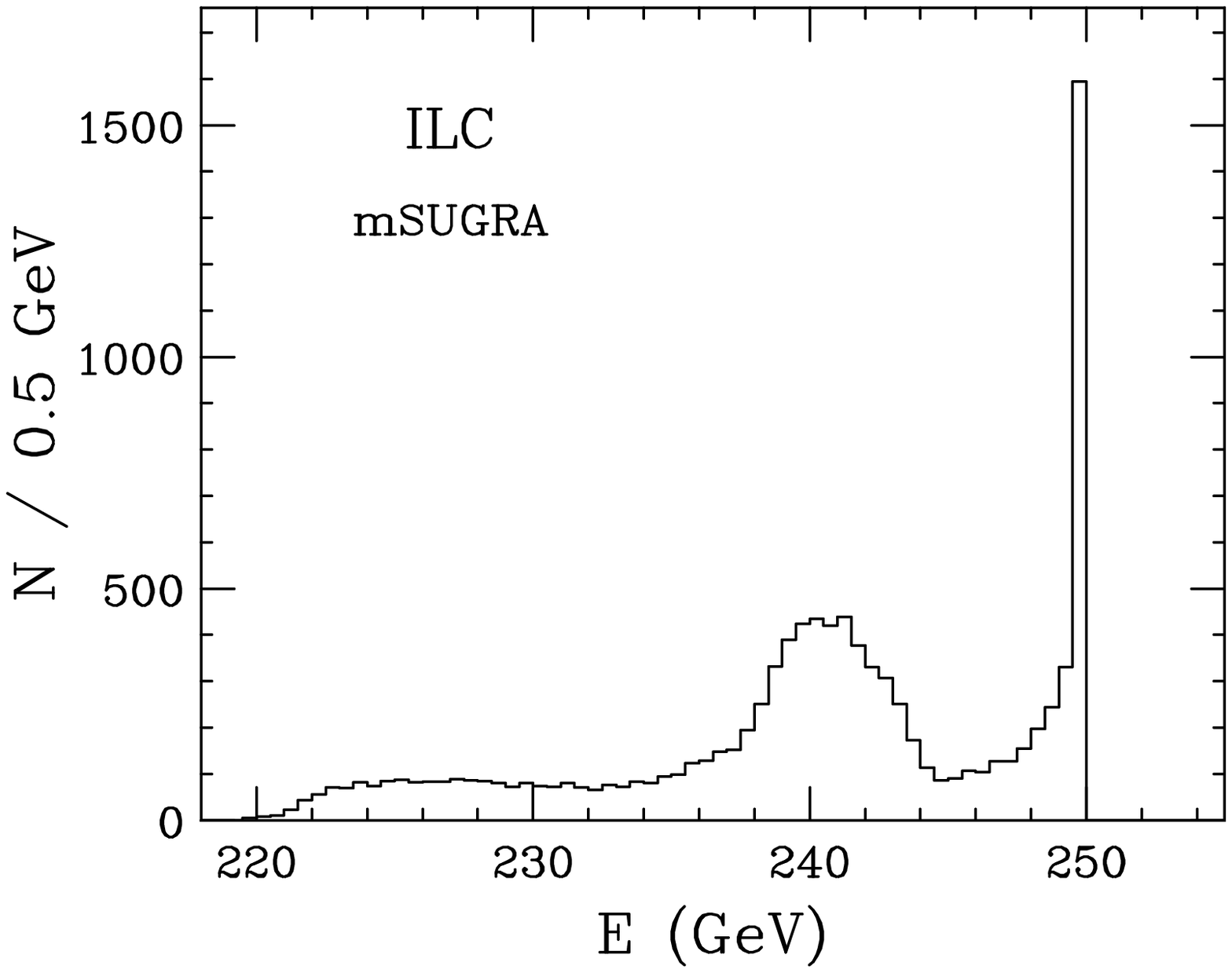} \qquad
\includegraphics{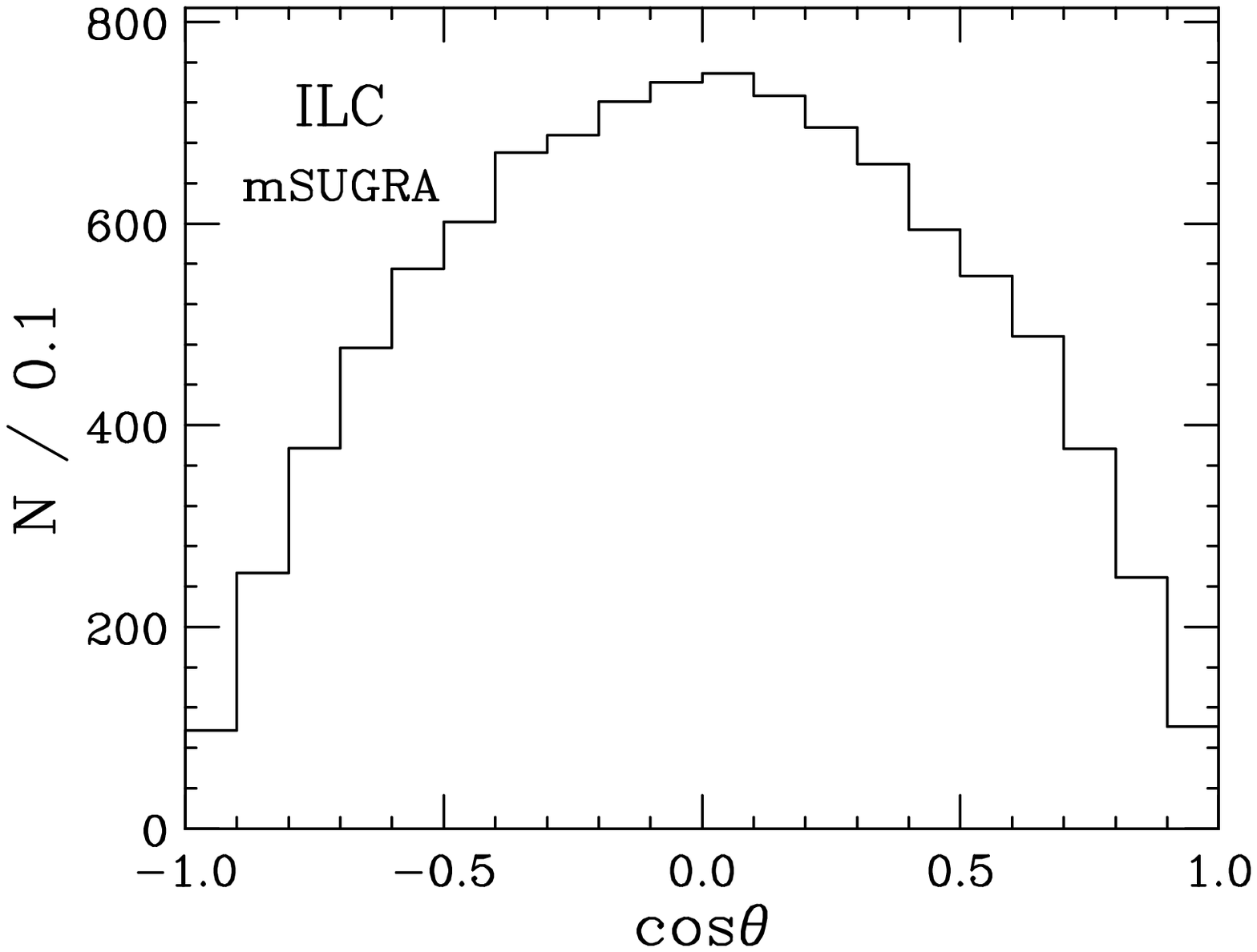}
}
\caption{Energy (left) and $\cos\theta$ (right) distributions for NLSP
  staus produced at the ILC in the mSUGRA model with $m_0 = 0$,
  $\mgaugino = 600~\gev$, $A_0 = 0$, $\tan\beta = 10$, $\mu >
  0$. Results are for $\sqrt{s} = 500~\gev$ and integrated luminosity
  $300~\ifb/\yr$.
\label{fig:distributions_mSUGRA} }
\end{figure}

The number of trapped staus per year for the mSUGRA model are given in
\figref{ILCresults_mSUGRA}.  The presence of additional accessible
superpartner states has a significant impact --- for all trap sizes
considered, large numbers of staus may be trapped even for beam
energies well above 475 GeV.  This is a consequence of the broad
energy distribution of NLSP staus, which in turn follows from the
existence of other fairly degenerate superpartners.  

\begin{figure}
\resizebox{3.25in}{!}{
\includegraphics{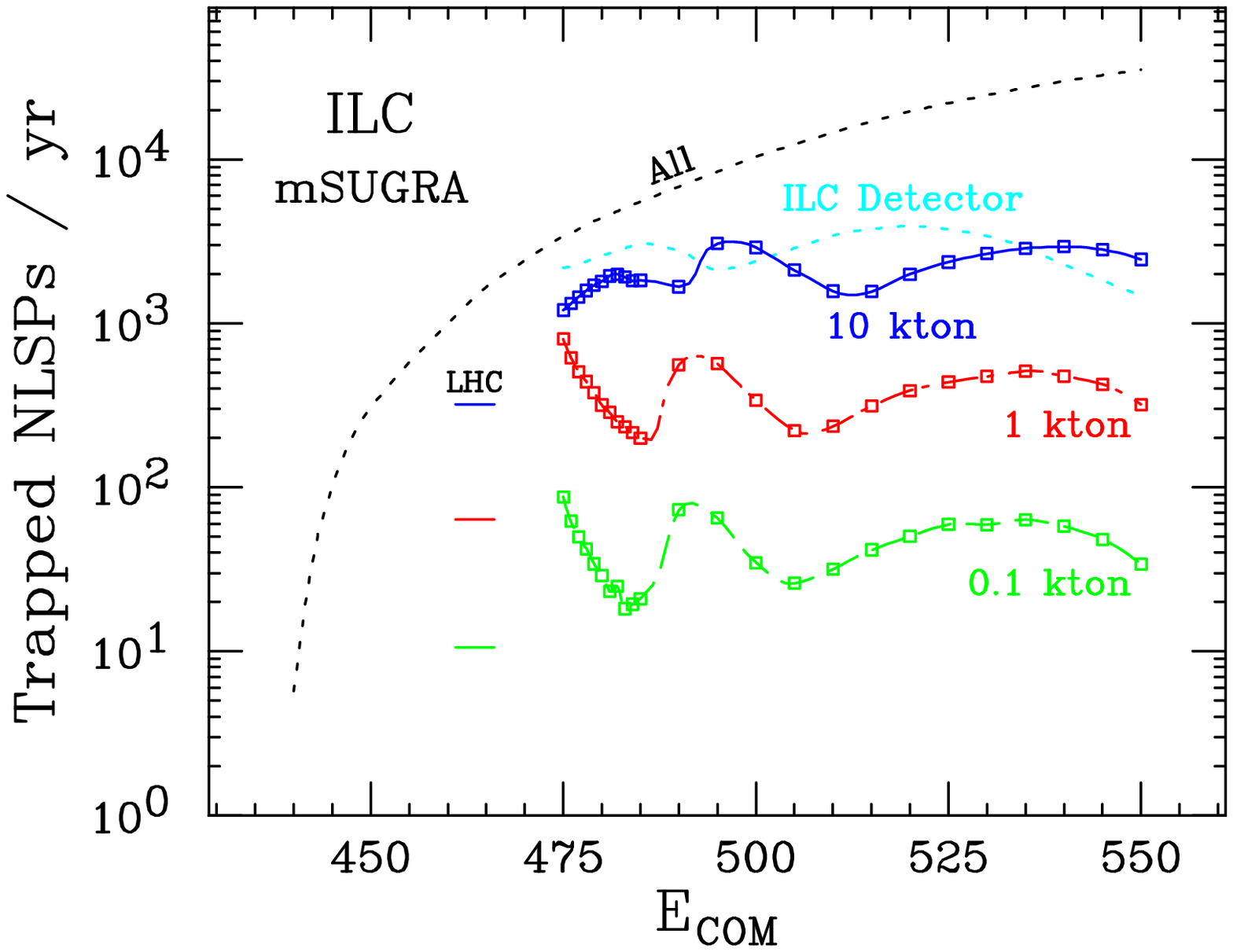} 
}
\caption{The number of sleptons trapped per year at the ILC in 10 kton
(solid), 1 kton (dot-dashed), and 0.1 kton (dashed) water traps.  The
total number of sleptons produced is also shown (upper dotted) along
with the number of sleptons trapped in the ILC detector (lower
dotted).  The trap shape and placement have been optimized, and we
assume $\rin = 10~\m$ and $\rinwe = 10~\mwe$, luminosity
$300~\ifb/\yr$ and the mSUGRA model with $m_0 = 0$, $\mgaugino =
600~\gev$, $A_0 = 0$, $\tan\beta = 10$, $\mu > 0$.  For comparison,
the number of trapped staus at the LHC for various trap volumes is
also shown.
\label{fig:ILCresults_mSUGRA} }
\end{figure}

In the mSUGRA case, we may also compare these ILC results directly
with results from the LHC analysis given above.  The LHC results for
$\mgaugino = 600~\gev$ from \secref{LHC} are given by the lines marked
``LHC'' in \figref{ILCresults_mSUGRA}.  Comparing results of
similarly-sized water traps, we find that for this particular model,
the ILC will be able to trap a factor of $\sim 10$ more staus than the
LHC. Again, for the reasons discussed above, a significant enhancement
of these ILC results may be possible if one considers inserting lead
between the ILC detector and the water trap, and as many as ${\cal
O}(10^4)$ staus may be trapped by reasonably sized water traps using
this more general approach.

\section{Implications and Conclusions}
\label{sec:implications}

Although the implications of supergravity for cosmology and particle
physics have been considered in great detail for decades, most work
has been centered on scenarios in which the LSP is a standard model
superpartner.  Here we have explored the gravitino LSP scenario.
Recent work has found significant cosmological motivations for this
possibility, as the gravitino may explain dark matter, and the
scenario may resolve current difficulties in Big Bang nucleosynthesis
and with leptogenesis.  We considered here a novel implication for
collider physics, namely that NLSP sleptons may be collected in water
traps before their decays to the gravitino.  These sleptons may then
be concentrated and transported to some quiet environment where their
decays may be studied in detail.

By optimizing the water trap shape and placement and considering a
variety of sizes, we have first explored the prospects for trapping
sleptons at the LHC.  The number that may be trapped is highly
model-dependent.  For minimal supergravity with $m_0= 0$, we find that
as many as $10^4$ staus may be stopped in a 10 kton trap when the
sleptons have mass around 100 GeV.  This is as light as is allowed by
current bounds.  For a less optimistic scenario, say, with 219 GeV
staus, hundreds and tens of sleptons may be trapped each year in 10
kton and 1 kton traps, respectively.

These results may be improved significantly if long-lived NLSP
sleptons are kinematically accessible at the ILC.  For the identical
case with 219 GeV sleptons discussed above, ${\cal O}(1000)$ sleptons
may be trapped each year in a 10 kton trap.  If only the NLSP is
accessible, this result may be achieved by tuning the beam energy so
that produced NLSPs barely escape the ILC detector.  The ability to
prepare initial states with well-known energies and the flexibility to
tune this energy are well-known advantages of the ILC.  Here, these
features are exploited in a qualitatively new way to produce slow
sleptons that are easily captured.

If there are additional superpartner states accessible at the ILC,
even tuning the beam energy is not necessary.  The cascade decays of
other superpartner states produce a broad distribution of slepton
energies, and so for a broad range of beam energies, some sleptons
will be captured in the trap. We have noted also that, by considering
the slightly more general possibility of placing lead or other dense
material between the ILC detector and the slepton trap, an order of
magnitude enhancement may be possible, allowing up to ${\cal O}(10^4)$
sleptons to be trapped per ILC year.

The analysis here is valid with minor revisions for traps composed of
any material.  For concreteness, however, we have considered traps
composed of water tanks, with the expectation that sleptons caught in
water will be easily concentrated and/or moved to quiet environments.

These results imply that high precision studies of slepton decays may
be possible.  There are many significant implications of such studies.
These have been considered in detail in
Refs.~\cite{Buchmuller:2004rq,Feng:2004gn}.  Briefly, simply by
counting the number of slepton decays as a function of time, the
slepton lifetime may be determined with high accuracy.  Given
thousands of sleptons, we expect a determination at the few percent
level.  The slepton decay width of \eqref{sfermionwidth} is a simple
function of the slepton and gravitino masses, and the slepton mass
will be constrained by analysis of the collider event kinematics, a
measurement of the slepton width therefore implies a high precision
measurement of the gravitino mass and, through \eqref{gravitinomass},
the supersymmetry breaking scale $F$.  Such measurements will provide
precision determinations of the relic density of superWIMP gravitino
dark matter, the contribution of supersymmetry breaking to vacuum
energy, and the opportunity for laboratory studies of late decay
phenomena relevant for Big Bang nucleosynthesis and the cosmic
microwave background.

The gravitino mass may also be determined, although not necessarily on
an event-by-event basis, by measuring the energy of slepton decay
products.  This provides a consistency check.  Alternatively, these
two methods, when combined, determine not only $m_{\gravitino}$, but
also the Planck mass $\mstar$.  This then provides a precision
measurement of Newton's constant on unprecedentedly small scales, and
the opportunity for a quantitative test of supergravity relations.

\begin{acknowledgments}
We are grateful to H.~Murayama for collaboration during the early
stages of this work and also to H.-C. Cheng, S.~Su and F.~Takayama for
valuable conversations.  We thank A.~Lankford, W.~Molzon, F.~Moortgat,
D.~Stoker, and especially D.~Casper for experimental insights.  The
work of JLF was supported in part by National Science Foundation
CAREER Grant PHY--0239817, and in part by the Alfred P.~Sloan
Foundation.

\vspace{.15in}

\noindent {\em Note added:}\ E-print
hep-ph/0409248~\cite{Hamaguchi:2004df}, in which long-lived sleptons
at colliders are also studied, appeared as this work was being
finalized. This paper studies the possibility of using an active
detector and also considers the idea of using $e^-e^-$ collisions at
the ILC.  The possibility of transporting trapped sleptons to a low
background environment, and the material-independent analysis of
optimizing trap shape and placement, both discussed at length here,
were not addressed.

\end{acknowledgments}

%%%%%%%%%%%%%%%%%%%%%%%%%%%%%%%%%%%%%%%%%%%%%%%%%%%%%%

\end{document}